\documentclass[12pt]{article}

\usepackage{graphics}
\usepackage{epsfig}
\usepackage{cite}
\input epsf

\title{Prompt photon and associated heavy quark production at hadron colliders with $k_T$-factorization}

\author{A.V.~Lipatov\footnote{lipatov@theory.sinp.msu.ru}, 
  M.A.~Malyshev\footnote{malyshev@theory.sinp.msu.ru}, N.P.~Zotov\footnote{zotov@theory.sinp.msu.ru}}

\begin{document}

\maketitle

\begin{center}

{\it Skobeltsyn Institute of Nuclear Physics,\\ 
Lomonosov Moscow State University,\\ 119991 Moscow, Russia}

\end{center}

\vspace{0.5cm}

\begin{center}

{\bf Abstract }

\end{center}

In the framework of the $k_T$-factorization approach, the production of prompt photons 
in association with a heavy (charm or beauty) quarks at high energies is studied. 
The consideration is based on the ${\cal O}(\alpha \alpha_s^2)$ off-shell amplitudes 
of gluon-gluon fusion and quark-(anti)quark interaction subprocesses.
The unintegrated parton densities in a proton are determined using the 
Kimber-Martin-Ryskin prescription. 
The analysis covers the total and differential cross sections 
and extends to specific angular correlations between the produced prompt photons and 
muons originating from the semileptonic decays of associated heavy quarks.
Theoretical uncertainties of our evaluations are studied and
comparison with the results of standard NLO pQCD calculations is performed.
Our numerical predictions are compared with the recent
experimental data taken by the D$\emptyset$ and CDF collaborations at the Tevatron.
Finally, we extend our results to LHC energies.

\vspace{1.0cm}

\noindent
PACS number(s): 12.38.Bx, 13.85.Qk

\newpage

Production of prompt photons\footnote{Usually the photons are called ”prompt” if they are
coupled to the interacting quarks.} in hadron-hadron collisions at high energies
is a subject of pointed theoretical and experimental investigations up to now\cite{1,2,3,4,5,6,7}.
Such processes have 
provided a direct probe of hard subprocess dynamics, since the produced 
photons are largely insensitive to the effects of final-state hadronization.
Inclusive prompt photon cross sections are strongly sensitive to the
parton content of a proton since, at leading order (LO) of 
Quantum Chromodynamics (QCD), such photons are produced mainly 
via quark-gluon Compton scattering or quark-antiquark
annihilation. An additional information about interaction can be extracted from the photon 
and associated jet production cross sections if flavor of the
produced jets is tagged. Such measurements have been performed recently
by the D$\emptyset$\cite{1,2} and CDF\cite{3,4} collaborations at the Tevatron
where $\gamma + c$-jet and $\gamma + b$-jet production 
cross sections as a function of photon transverse momentum $p_T^\gamma$
have been reported. The perturbative QCD predictions\cite{7} calculated at next-to-leading order (NLO) 
agree reasonably well with the measured cross sections at relatively low $p_T^\gamma$ values, 
up to $p_T^\gamma \sim 70$~GeV. 
However, the substantial disagreement between theory and data for both 
$\gamma + b$-jet and $\gamma + c$-jet production at large $p_T^\gamma$
was observed\cite{1,2,3}. In the present note we will analyse recent D$\emptyset$\cite{1,2} and CDF\cite{3,4} data
as well as previous CDF measurements\cite{5,6} using the $k_T$-factorization approach of QCD\cite{8}.
This approach has been successfully applied by us, in particular, to 
heavy flavour production\cite{9} and inclusive 
prompt photon production\cite{10,11,12,13}
at the HERA, Tevatron and LHC energies\footnote{A detailed description and discussion 
of the $k_T$-factorization approach can be found in\cite{14}.}.
An additional motivation for our study is that such processes are background processes for the physics
beyound the Standard Model (SM), for example the production of 
excited quarks or gauge-mediated supersymmetry breaking with 
neutralinos radiatively decaying to gravitinos\cite{15}, and, therefore, 
it is necessary to have a realistic estimations of corresponding cross sections within QCD.

First application of $k_T$-factorization approach to production of photons associated with the charm 
or beauty quarks have been performed in our previous paper\cite{16}. The consideration 
was based on the ${\cal O}(\alpha \alpha_s^2)$ amplitude for the production of a single photon associated with a 
quark pair in the fusion of two off-shell gluons $g^*g^* \to \gamma Q\bar Q$ (see Fig.~1a).
A reasonably good agreement between the  
numerical predictions and the Tevatron data\cite{5,6} was obtained in the 
region of relatively low $p_T^\gamma$ where off-shell gluon fusion dominates.
However, the quark-induced subprocesses become more important at moderate and large $p_T^\gamma$
and therefore should be taken into account.
In the present note we extend our previous predictions\cite{15} by including into 
the consideration two additional ${\cal O}(\alpha \alpha_s^2)$ subprocesses, 
namely $q^* \bar q^* \to \gamma Q \bar Q$ (Fig.~1b) and
$q^*(\bar q^*)Q \to \gamma q(\bar q)Q$ (Fig.~1c), where $Q$ is the charm or beauty quark. 
These subprocesses give a sizeble contribution in the $p_T^\gamma$ region probed by
measurements\cite{1,2,3,4}.
In other aspects we follow the 
approach described in\cite{16}. 
Our main goal is to give a systematic analysis of available D$\emptyset$ and CDF data\cite{1,2,3,4,5,6} 
in the framework of $k_T$-factorization. Specially we study different sources of theoretical uncertainties. 
Such calculations are performed for the first time.

Let us start from a very short review of calculation steps. We describe first the evaluation
of the off-shell $q^* \bar q^* \to \gamma Q \bar Q$ and $q^*(\bar q^*)Q \to \gamma q(\bar q)Q$ production amplitudes.
We denote the 4-momenta of incoming 
quarks, produced photon and outgoing quarks by $k_1$, $k_2$, $p$, $p_1$ and $p_2$, respectively; 
and polarization 4-vector of produced prompt photon is given by $\epsilon$.
The corresponding partonic amplitudes can be written as follows:
$$
  \displaystyle {\cal M}(q^* \bar q^* \to \gamma Q \bar Q) = e e_q g^2 t^a \delta^{ab} t^b {1\over (p_1 + p_2)^2}
  \displaystyle \epsilon_{\lambda}(p) L_{q\bar q}^{(1)\,\mu \lambda} L_{q\bar q}^{(1)\,\nu} g_{\mu \nu} + \atop {
  \displaystyle + e e_Q g^2 t^a \delta^{ab} t^b {1\over (k_1 + k_2)^2} \epsilon_{\lambda}(p) L_{q\bar q}^{(2)\,\mu} 
  \displaystyle L_{q\bar q}^{(2)\,\nu \lambda} g_{\mu \nu} }, \eqno(1)
$$
$$
  \displaystyle {\cal M}(q^* Q \to \gamma q Q) = e e_q g^2 t^a \delta^{ab} t^b {1\over (k_2 - p_2)^2}
  \displaystyle \epsilon_{\lambda}(p) L_{qQ}^{(1)\,\mu \lambda} L_{qQ}^{(1)\,\nu} g_{\mu \nu} + \atop {
  \displaystyle + e e_Q g^2 t^a \delta^{ab} t^b {1\over (k_1 - p_1)^2} \epsilon_{\lambda}(p) L_{qQ}^{(2)\,\mu} 
  \displaystyle L_{qQ}^{(2)\,\nu \lambda} g_{\mu \nu} }, \eqno(2)
$$
\noindent
where
$$
  L_{q\bar q}^{(1)\,\mu \lambda} = \bar u(k_2) \left[ \gamma^\mu {\hat k_1 - \hat p + m_q \over (k_1 - p)^2 - m_q^2}
  \gamma^\lambda + \gamma^\lambda { - \hat k_2 + \hat p + m_q\over ( - k_2 + p)^2 - m_q^2} \gamma^\mu \right] u(k_1), \eqno(3)
$$
$$
  L_{q\bar q}^{(1)\,\nu} = \bar u(p_1) \gamma^\nu u(p_2), \eqno(4)
$$
$$
  L_{q\bar q}^{(2)\,\nu \lambda} = \bar u(p_1) \left[ \gamma^\nu { - \hat p_2 - \hat p + m_Q \over ( - p_2 - p)^2 - m_Q^2}
  \gamma^\lambda + \gamma^\lambda { \hat p_1 + \hat p + m_Q\over (p_1 + p)^2 - m_Q^2} \gamma^\nu \right] u(p_2), \eqno(5)
$$
$$
  L_{q\bar q}^{(2)\,\mu} = \bar u(k_2) \gamma^\mu u(k_1), \eqno(6)
$$
$$
  L_{qQ}^{(1)\,\mu \lambda} = \bar u(p_1) \left[ \gamma^\lambda {\hat p_1 + \hat p + m_q \over (p_1 + p)^2 - m_q^2}
  \gamma^\mu + \gamma^\mu { \hat k_1 - \hat p + m_q\over (k_1 - p)^2 - m_q^2} \gamma^\lambda \right] u(k_1), \eqno(7)
$$
$$
  L_{qQ}^{(1)\,\nu} = \bar u(p_2) \gamma^\nu u(k_2), \eqno(8)
$$
$$
  L_{qQ}^{(2)\,\nu \lambda} = \bar u(p_2) \left[ \gamma^\lambda {\hat p_2 + \hat p + m_Q \over (p_2 + p)^2 - m_Q^2}
  \gamma^\nu + \gamma^\nu { \hat k_2 - \hat p + m_Q\over (k_2 - p)^2 - m_Q^2} \gamma^\lambda \right] u(k_2), \eqno(9)
$$
$$
  L_{qQ}^{(2)\,\mu} = \bar u(p_1) \gamma^\mu u(k_1). \eqno(10)
$$
\noindent 
Here $e$, $e_q$ and $e_Q$ are the electron, light and heavy quark (fractional) electric charges,
$g$ is the strong charge, $m_q$ and $m_Q$ are the light and heavy quark masses, $a$ and $b$ are 
the eight-fold color indexes.

When we calculate the matrix element squared, the summation over the produced photon
polarizations is carried with $\sum \epsilon^\mu(p) \epsilon^{*\,\nu}(p) = - g^{\mu \nu}$,
and the spin density matrix for all on-shell spinors in final state is taken in the usual
form $\sum u(p) \bar u(p) = \hat p + m$.
However, in the case of initial off-shell quarks on-shell quark spin density matrix 
has to be replaced by a more complicated expression\footnote{We neglected virtualities 
of incoming heavy quarks in $q^*(\bar q)Q \to \gamma q (\bar q)Q$ amplitude
and treat them as on-shell.}.
To evaluate it we follow a simple approximation proposed in\cite{10}. We ”extend” the original 
diagram and consider the off-shell quark line as internal
line in the extended diagram. The ”extended” process looks like follows: the initial on-shell
quark with 4-momentum $p$ and mass $m$ radiates a quantum (say, photon or gluon) and
becomes an off-shell quark with 4-momentum $k$. So, for the extended diagram squared
we can write:
$$
  |{\cal M}|^2 \sim {\rm tr} \left[ \bar {\cal T}^\mu {\hat k + m\over k^2 - m^2} \gamma^\nu\, u(p) \bar u(p) \, 
  \gamma_\nu {\hat k + m\over k^2 - m^2} {\cal T}_\mu \right], \eqno(11)
$$
\noindent 
where $\cal T$ is the rest of the original matrix element.
The expression presented between $\bar {\cal T}^\mu$ and ${\cal T}_\mu$ now plays the role 
of the off-shell quark spin density matrix. Using the standard on-shell condition 
$\sum u(p)\bar u(p) = \hat p + m$ and performing the Dirac algebra one obtains
in the massless limit $m \to 0$:
$$
  |{\cal M}|^2 \sim {2\over (k^2)^2} {\rm tr} \left[ \bar {\cal T}^\mu \left( k^2 \hat p - 
  2 (p \cdot k) \hat k\right) {\cal T}_\mu \right]. \eqno(12)
$$
\noindent 
Now, using the Sudakov decomposition $k = x p + k_T$ (where $k_T$ is the 
off-shell quark non-zero transverse 4-momentum, $k^2 = k_T^2 = - {\mathbf k}_T^2$) 
and neglecting the second term in the
parentheses in~(12) in the small-$x$ limit, we easily obtain:
$$
  |{\cal M}|^2 \sim {2\over x k^2} {\rm tr} \left[ \bar {\cal T}^\mu x \hat p {\cal T}_\mu \right]. \eqno(13)
$$
\noindent 
Essentially, we have neglected here the negative light-cone momentum fraction of the incoming quark. 
The properly normalized off-shell spin density matrix is given by $x \hat p$, while
the factor $2/x k^2$ has to be attributed to the quark distribution function (determining its
leading behavior). With this normalization, we successfully recover the on-shell collinear 
limit when $k$ is collinear with $p$. 

Further calculations are straighforward and in other respects follow the standard QCD Feynman
rules\footnote{We neglected the contributions from the so-called
fragmentation mechanisms. It is because after applying the isolation cut (see\cite{1,2,3,4,5,6}) 
these contributions amount only to about 10\% of the visible cross section. 
The isolation requirement and additional conditions
which preserve our calculations from divergences have been specially discussed in\cite{12}.}. 
The evaluation of traces was done using the algebraic 
manipulation system {\textsc form}\cite{17}. We do not list here the obvious expressions 
because of lack of space. The analytic expression for the $|\bar {\cal M}|^2(g^*g^* \to \gamma Q\bar Q)$ 
has been derived
in our previous paper\cite{12}. 

According to the $k_T$-factorization theorem, to calculate the cross section of any process
one should convolute the off-shell partonic cross sections 
with the corresponding unintegrated (dependent from the transverse momenta) quark and/or gluon distributions in a proton:
$$
  \displaystyle \sigma = \sum_{i,j = q,\,g} \int {\hat \sigma}_{ij}^*(x_1, x_2, {\mathbf k}_{1T}^2, {\mathbf k}_{2T}^2) \, 
  \displaystyle f_i(x_1,{\mathbf k}_{1T}^2,\mu^2) f_j(x_2,{\mathbf k}_{2T}^2,\mu^2) \times \atop { 
  \displaystyle \times dx_1 dx_2 \, d{\mathbf k}_{1T}^2 d{\mathbf k}_{2T}^2 {d\phi_1\over 2\pi} {d\phi_2\over 2\pi} }, 
  \eqno(14)
$$

\noindent
where ${\hat \sigma}_{ij}^*(x_1, x_2, {\mathbf k}_{1T}^2, {\mathbf k}_{2T}^2) \sim |\bar {\cal M}^2_{ij}({\mathbf k}_{1T}^2, {\mathbf k}_{2T}^2)| \otimes d\Phi$
is the relevant off-shell partonic cross section and incoming partons have fractions $x_1$ and $x_2$ of 
initial protons longitudinal momenta, transverse momenta 
${\mathbf k}_{1T}$ and ${\mathbf k}_{2T}$ and azimuthal angles $\phi_1$ and $\phi_2$.
The multiparticle phase space $d\Phi = \Pi d^3 p_i / 2 E_i \delta^{(4)} (\sum p^{\rm in} - \sum p^{\rm out} )$
is parametrized in terms of transverse momenta, rapidities and azimuthal angles
of relevant particles:
$$
  { d^3 p_i \over 2 E_i} = {\pi \over 2} \, d {\mathbf p}_{iT}^2 \, dy_i \, { d \psi_i \over 2 \pi}. \eqno(15)
$$
From~(14) and~(15) we can obtain following expressions:
$$
  \displaystyle \sigma_{gg} = \int {1\over 256\pi^3 (x_1 x_2 s)^2} |\bar {\cal M}(g^* g^* \to \gamma Q \bar Q)|^2 \times \atop 
  \displaystyle \times f_g(x_1,{\mathbf k}_{1 T}^2,\mu^2) f_g(x_2,{\mathbf k}_{2 T}^2,\mu^2) 
  d{\mathbf k}_{1 T}^2 d{\mathbf k}_{2 T}^2 d{\mathbf p}_{1 T}^2 {\mathbf p}_{2 T}^2 dy dy_1 dy_2 
  {d\phi_1\over 2\pi} {d\phi_2\over 2\pi} {d\psi_1\over 2\pi} {d\psi_2\over 2\pi}, \eqno(16)
$$
$$
  \displaystyle \sigma_{q\bar q} = \sum_q \int {1\over 256\pi^3 (x_1 x_2 s)^2} |\bar {\cal M}(q^* \bar q^* \to \gamma Q \bar Q)|^2 \times \atop 
  \displaystyle \times f_q(x_1,{\mathbf k}_{1 T}^2,\mu^2) f_q(x_2,{\mathbf k}_{2 T}^2,\mu^2) 
  d{\mathbf k}_{1 T}^2 d{\mathbf k}_{2 T}^2 d{\mathbf p}_{1 T}^2 {\mathbf p}_{2 T}^2 dy dy_1 dy_2 
  {d\phi_1\over 2\pi} {d\phi_2\over 2\pi} {d\psi_1\over 2\pi} {d\psi_2\over 2\pi}, \eqno(17)
$$
\noindent
where $y$ is the rapidity of the produced prompt photon, 
${\mathbf p}_{1T}$, ${\mathbf p}_{2T}$, $y_1$, $y_2$, $\psi_1$ and $\psi_2$ are the 
transverse momenta, rapidities and azimuthal angles of final state quarks, respectively.
Similar expression can be easily written for $q^* (\bar q^*) Q \to \gamma q (\bar q) Q$ contribution.
If we average~(16) and~(17) over $\phi_{1}$ and $\phi_{2}$ 
and take the limit ${\mathbf k}_{1 T}^2 \to 0$ and ${\mathbf k}_{2 T}^2 \to 0$,
then we recover corresponding expressions of usual collinear QCD approximation.

Concerning the unintegrated quark and gluon densities in 
a proton, we apply the Kimber-Martin-Ryskin (KMR) approach\cite{18} to calculate them. 
The KMR approach is the formalism
to construct the unintegrated parton distributions from the known conventional ones. 
In this case the unintegrated quark and gluon distributions are given by
$$
  \displaystyle f_q(x,{\mathbf k}_T^2,\mu^2) = T_q({\mathbf k}_T^2,\mu^2) {\alpha_s({\mathbf k}_T^2)\over 2\pi} \times \atop {
  \displaystyle \times \int\limits_x^1 dz \left[P_{qq}(z) {x\over z} q\left({x\over z},{\mathbf k}_T^2\right) \Theta\left(\Delta - z\right) + P_{qg}(z) {x\over z} g\left({x\over z},{\mathbf k}_T^2\right) \right],} \eqno (18)
$$
$$
  \displaystyle f_g(x,{\mathbf k}_T^2,\mu^2) = T_g({\mathbf k}_T^2,\mu^2) {\alpha_s({\mathbf k}_T^2)\over 2\pi} \times \atop {
  \displaystyle \times \int\limits_x^1 dz \left[\sum_q P_{gq}(z) {x\over z} q\left({x\over z},{\mathbf k}_T^2\right) + P_{gg}(z) {x\over z} g\left({x\over z},{\mathbf k}_T^2\right)\Theta\left(\Delta - z\right) \right],} \eqno (19)
$$
\noindent
where $P_{ab}(z)$ are the usual unregulated LO DGLAP splitting 
functions. The theta functions which appears
in~(18) and~(19) imply the angular-ordering constraint $\Delta = \mu/(\mu + |{\mathbf k}_T|)$ 
specifically to the last evolution step to regulate the soft gluon
singularities. 
Numerically, for the input we have used LO parton densities $xq(x,\mu^2)$ and 
$xg(x,\mu^2)$ from recent MSTW'2008 set\cite{19}.

Other essential parameters were taken as follows:
renormalization and factorization scales $\mu = \xi p_T^\gamma$ (where
we vary the parameter $\xi$ between 1/2 and 2 about the default value $\xi = 1$
in order to estimate the scale uncertainties of our calculations),
LO formula for the strong coupling constant $\alpha_s(\mu^2)$ 
with $n_f = 4$ and $\Lambda_{\rm QCD} = 200$ MeV, 
such that $\alpha_s(M_Z^2) = 0.1232$.
We set the charm and beauty quark masses 
to $m_c = 1.5$~GeV and $m_b = 4.75$~GeV. 
The multidimensional integration has been performed
by means of the Monte Carlo technique, using the routine \textsc{vegas}\cite{20}.
The full C$++$ code is available from the authors on 
request\footnote{lipatov@theory.sinp.msu.ru}.

We now are in a position to present our predictions.
In Figs.~2 --- 5 we confront the calculated $\gamma + b$-jet 
differential cross sections as a function of photon transverse 
momentum with the recent data\cite{1,2,3,4} taken by the D$\emptyset$ and CDF collaborations at $\sqrt s = 1960$~GeV.
These data refer to the central kinematic region for heavy quark jet,
and the D$\emptyset$ measurements\cite{2} have been performed in two regions of kinematics, 
defined by $y^\gamma y^{\rm jet} > 0$ and $y^\gamma y^{\rm jet} < 0$.
For comparison we also plot the NLO QCD predictions\cite{7} taken from\cite{1,2,3,4}.
We find that the full set of experimental data are reasonably well described by
the $k_T$-factorization approach. One can see that the shape and absolute normalization of measured 
cross sections are adequately reproduced.
The overall agreement of 
our predictions and the data 
at moderate $p_T^\gamma$ is on the same level as it was achieved
in the framework of NLO QCD approximation, and the difference appears at small and 
large $p_T^\gamma$ (see Figs.~2, 4 and~5). 
However, in the case of $\gamma + c$-jet production, the situation is a bit worse:
we find a substantial disagreement between our predictions and D$\emptyset$ data at high 
$p_T^\gamma$ (see Fig.~6). 
As it was noted in\cite{2},  
the disagreement can be reduced if, in particular, additional 
contributions involving intrinsic charm\cite{21} are taken into account. 
The consideration of such contributions is out of our paper.
Note, however, that very recent CDF data\cite{3} for
$\gamma +c$-jet production 
are well described by the
$k_T$-factorization in a whole $p_T$ range (see Fig.~7).

The relative contributions of $g^* g^* \to \gamma Q\bar Q$ and 
$q^*(\bar q^*)Q \to \gamma q (\bar q) Q$ are approximately equal to each other in the
considered kinematical region. These subprocesses give a main contribution 
to the cross section up to $p_T^\gamma \simeq 60$~GeV.
At higher photon transverse momenta the $q^* {\bar q}^* \to \gamma Q\bar Q$ subprocess dominates.
The scale uncertainties of our calculations at high and low $p_T^\gamma$ 
are defined by uncertainties of corresponding leading production mechanisms and 
connected, in particular, with the unintegrated gluon and quark densities at low and high $x$.

The CDF collaboration have reported data\cite{5,6} which 
refer to the muons which originate from the
semileptonic decays of associated charmed or beauty quarks. These 
measurements have been performed in the kinematic region defined by
$|\eta^\gamma| < 0.9$, $p_T^\mu > 4$~GeV and $|\eta^\mu| < 1$
at $\sqrt s = 1800$~GeV. To produce muons from charmed and beauty quarks,
we first convert them into a $D$ or $B$ hadrons using
the Peterson fragmentation function\cite{22} and then 
simulate their semileptonic decay according to the
standard electroweak theory. Additionally, the cascade decays $b\to c\to \mu$
have been taken into account. We set the fragmentation 
parameters $\epsilon_c = 0.06$ and $\epsilon_b = 0.006$ and
corresponding branching fractions to
$f(c \to \mu) = 0.0969$, $f(b \to \mu) = 0.1071$ and 
$f(b\to c \to \mu) = 0.0802$\cite{23}.
Our predictions for $\gamma + \mu$ cross sections as a functions 
of transverse momentum $p_T^\gamma$ and azimuthal angle difference 
$\Delta \phi$ (i.e. difference between the produced photon and muon transverse momenta) are shown in Figs.~8 and~9.
We find that the $k_T$-factorization predictions 
slightly overestimate
the measured transverse momentum distributions but agree with data within the uncertainties.

Obtained perfect description of the $\Delta \phi$ distribution is notable. 
The important role of such observables for understanding 
an interaction dynamics is well known\footnote{See, for example,\cite{14}.}.
In particular, they give an additional insight into the 
effective contributions from higher-order QCD processes.
In the naive LO QCD approximation, the distribution over $\Delta \phi$
must be simply a delta function $\delta(\Delta \phi - \pi)$ since 
the produced particles are back-to-back in the transverse plane and 
are balanced in $p_T$ due to momentum conservation.
When higher-order QCD processes are considered, the
presence of additional quarks and/or gluons in the final state
allows the $\Delta \phi$ distribution to be more spread.
In the $k_T$-factorization approach, taking into 
account the non-vanishing initial parton
transverse momentum ${\mathbf k}_{T}$ leads to 
the violation of back-to-back kinematics even at leading order.
Despite the fact that using the $2 \to 3$ matrix elements 
in our consideration instead the $2 \to 2$ ones makes the difference between the 
$k_T$-factorization predictions and the collinear approximation of QCD 
not well pronounced, the CDF data clearly favour the $k_T$-factorization results.

Now we can try to extend our predictions to LHC energies. 
As a representative example, we define the kinematical region 
by the requirements
$|y^\gamma| < 2.5$, $25 < p_T^\gamma < 400$~GeV, $|y^{\rm jet}| < 2.2$ and $18 < p_T^{\rm jet} < 200$~GeV.
Such region is coincide with one defined in recent analyses performed by the 
CMS and ATLAS collaborations\cite{24,25}.
Our predictions for differential $\gamma +b$-jet cross sections as a function 
of photon transverse momentum $p_T^\gamma$ and rapidity $y^\gamma$ are shown in Fig.~10.
The corresponding experimental data at $\sqrt s = 7$~TeV are still waited.

To conclude, in the present note we apply the $k_T$-factorization approach 
to the analysis of recent CDF and D0 experimental data on the associated photon and 
heavy quark production.
Using the off-shell ($k_T$-dependent) matrix elements of gluon-gluon fusion and quark-(anti)quark
interaction subprocesses, we obtained a good agreement between our 
predictions and the Tevatron data. 
As it was noted in\cite{1}, our results agree better with the Tevatron data 
than the NLO QCD ones.
It is important for further studies of small-$x$ physics, 
and, in particular, for searches of effects of new physics beyond the SM at modern 
hadron colliders.

{\sl Acknowledgements.} The authors would like to thank 
S.P.~Baranov and H.~Jung for their encouraging interest 
and helpful discussions. We are very grateful to 
D.~Bandurin for fruitful discussions of D$\emptyset$ measurements and
K.~Vellidis and T.~Yang for providing us with latest CDF experimental data.
We are also grateful to DESY Directorate for the support in the 
framework of Moscow --- DESY project on Monte-Carlo
implementation for HERA --- LHC.
A.V.L. and M.A.M. was supported in part by the grant of president of 
Russian Federation (MK-3977.2011.2).
This research was supported by the 
FASI of Russian Federation (grant NS-3920.2012.2),
FASI state contract 02.740.11.0244, 
RFBR grant 11-02-01454-a and the RMES (grant the Scientific Research on High Energy Physics).

\newpage

\begin{figure}
\begin{center}
\epsfig{figure=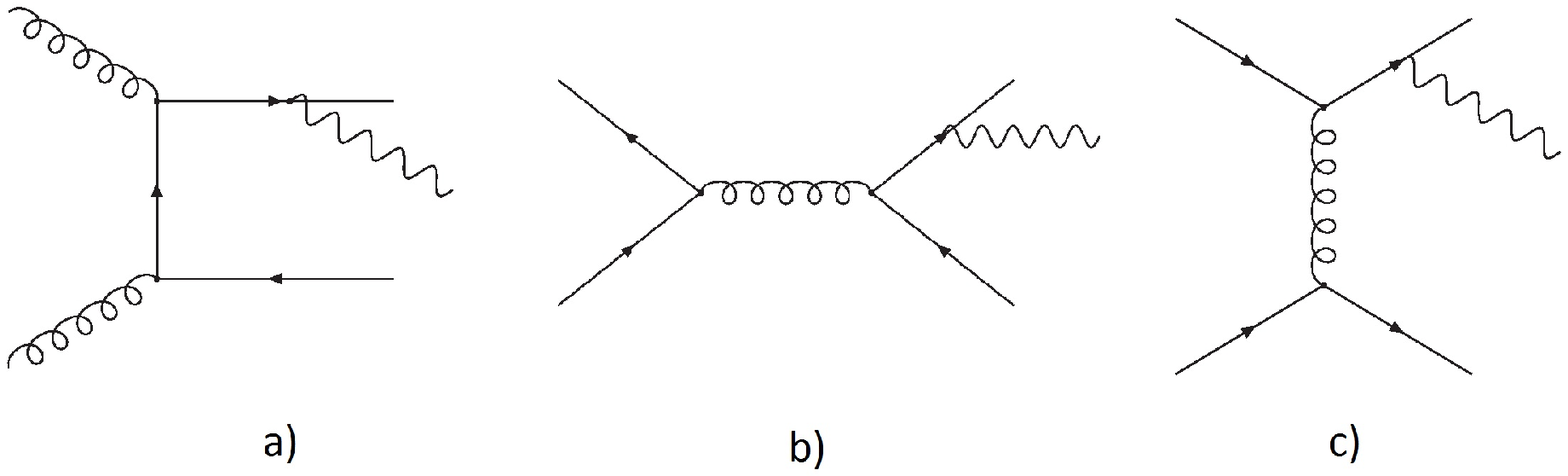, width = 10cm}
\caption{Examples of Feynman diagrams contributing to associated photon and heavy quark production. Full 
set of diagrams can be obtained by permutations of quark and photon lines.}
\label{fig1}
\end{center}
\end{figure}

\begin{figure}
\begin{center}
\epsfig{figure=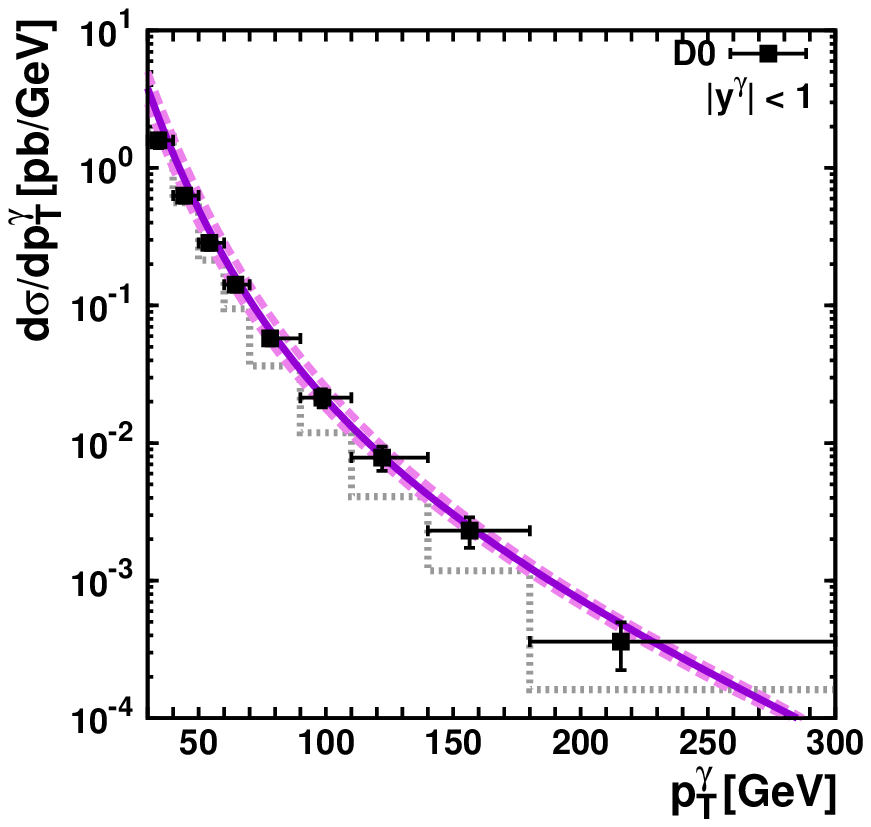, width = 8.1cm}
\epsfig{figure=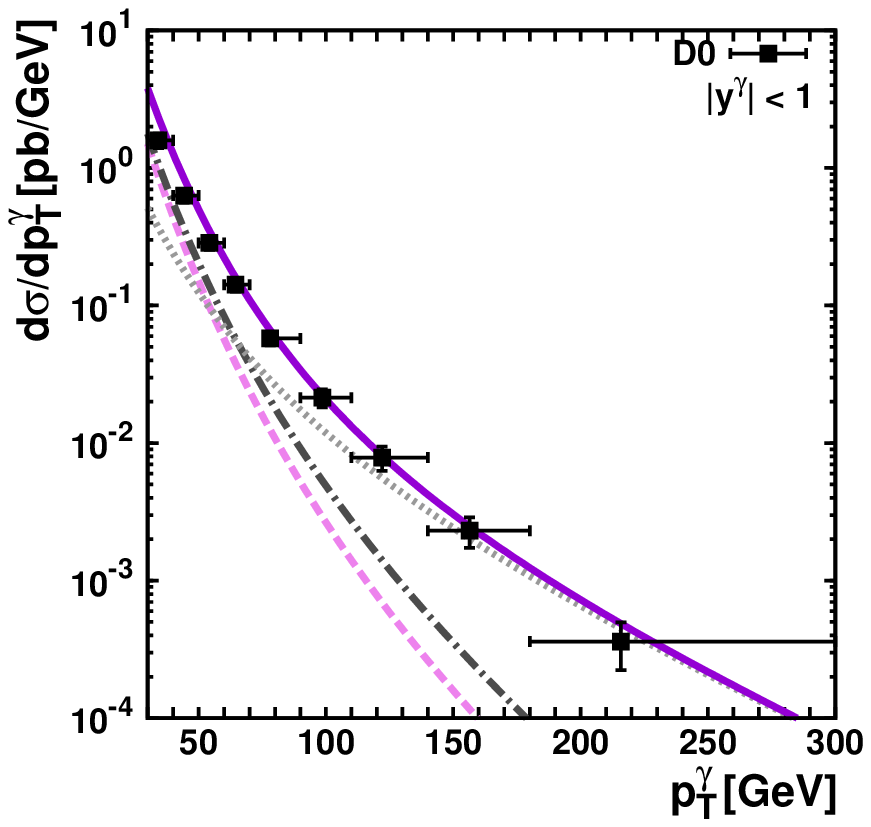, width = 8.1cm}
\epsfig{figure=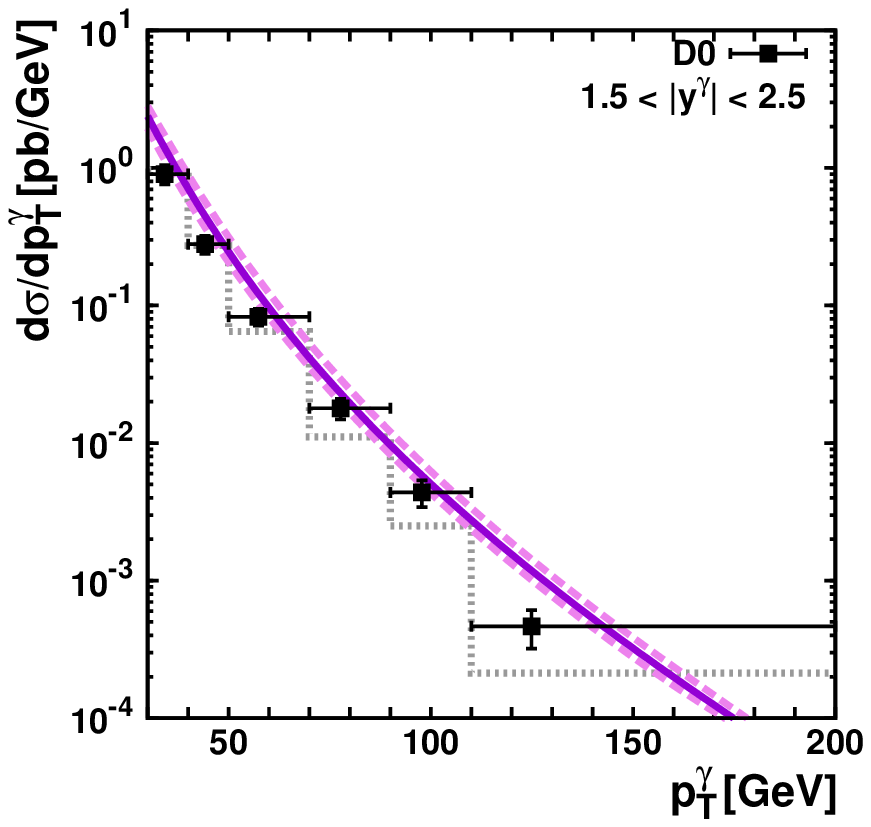, width = 8.1cm}
\epsfig{figure=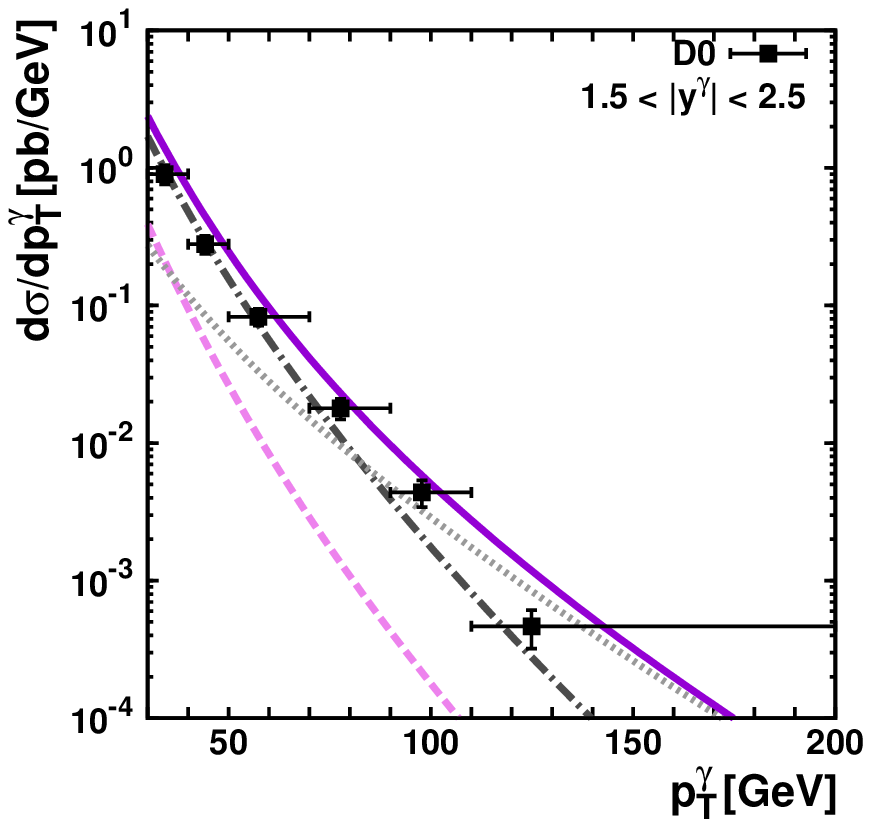, width = 8.1cm}
\caption{The associated $\gamma + b$-jet cross section
as a function of photon transverse momentum $p_T^\gamma$ in the kinematical region defined by
$|y^{\rm jet}| < 1.5$ and $p_T^{\rm jet} > 15$~GeV at $\sqrt s = 1960$~GeV. 
Left panels: the solid curve corresponds to the KMR predictions at the 
default scale $\mu = E_T^\gamma$, whereas the upper and 
lower dashed curves correspond to scale variations described in the text.
The dotted histogram represents the NLO pQCD predictions\cite{7} listed in\cite{1}.
Right panels: the different contributions to the $\gamma + b$-jet cross section. The dashed,
dotted and dash-dotted curves correspond to the contributions from the $g^*g^* \to \gamma Q\bar Q$, 
$q^*\bar q^* \to \gamma Q\bar Q$ and $q^*(\bar q^*)Q \to \gamma q (\bar q) Q$
subprocesses, respectively. The solid curve represents their sum. 
The experimental data are from D$\emptyset$\cite{1}. }
\label{fig2}
\end{center}
\end{figure}

\begin{figure}
\begin{center}
\epsfig{figure=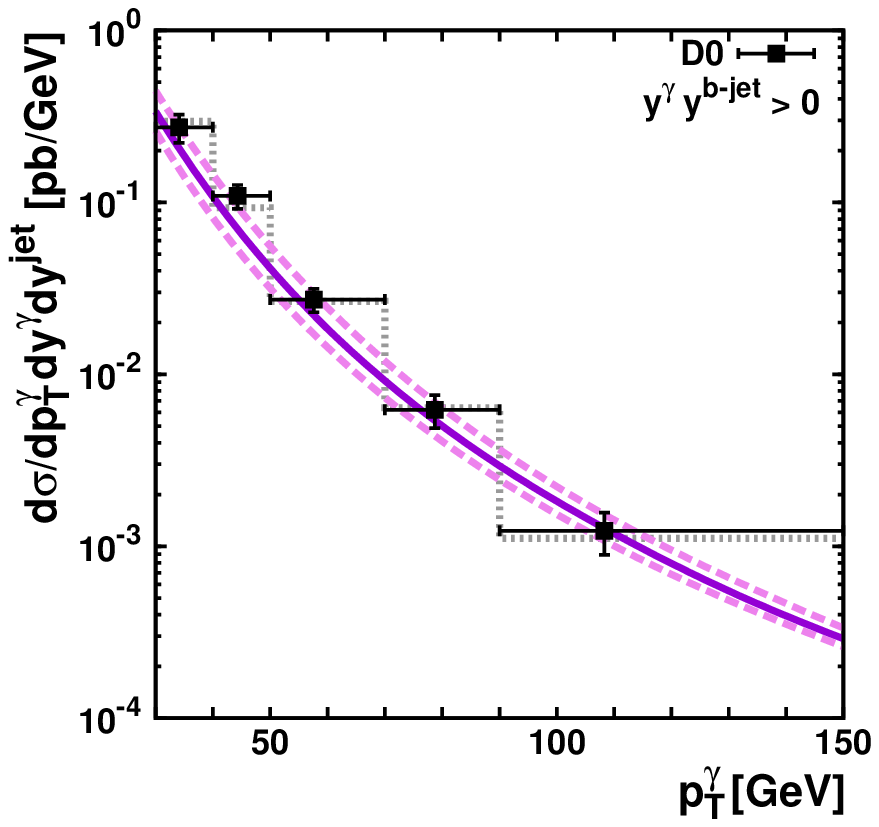, width = 8.1cm}
\epsfig{figure=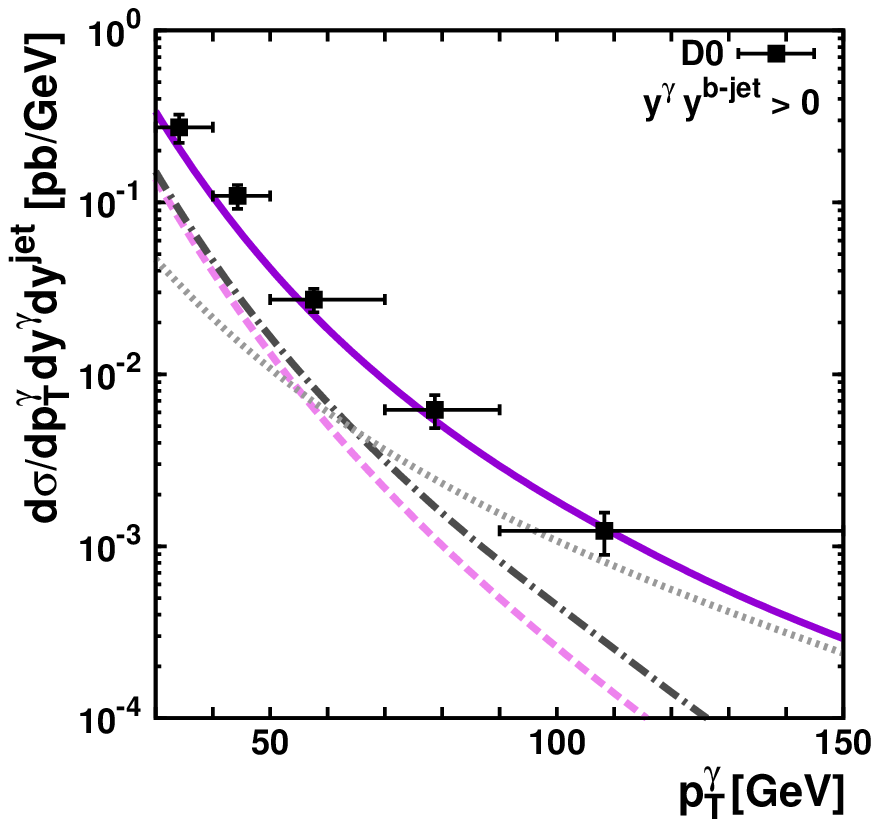, width = 8.1cm}
\epsfig{figure=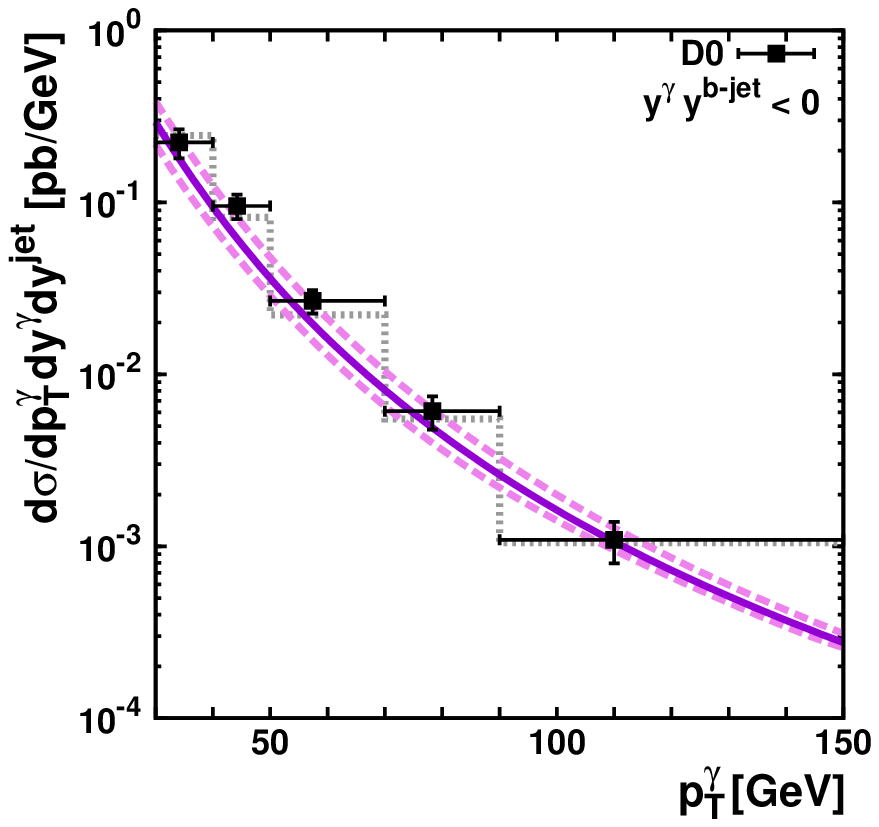, width = 8.1cm}
\epsfig{figure=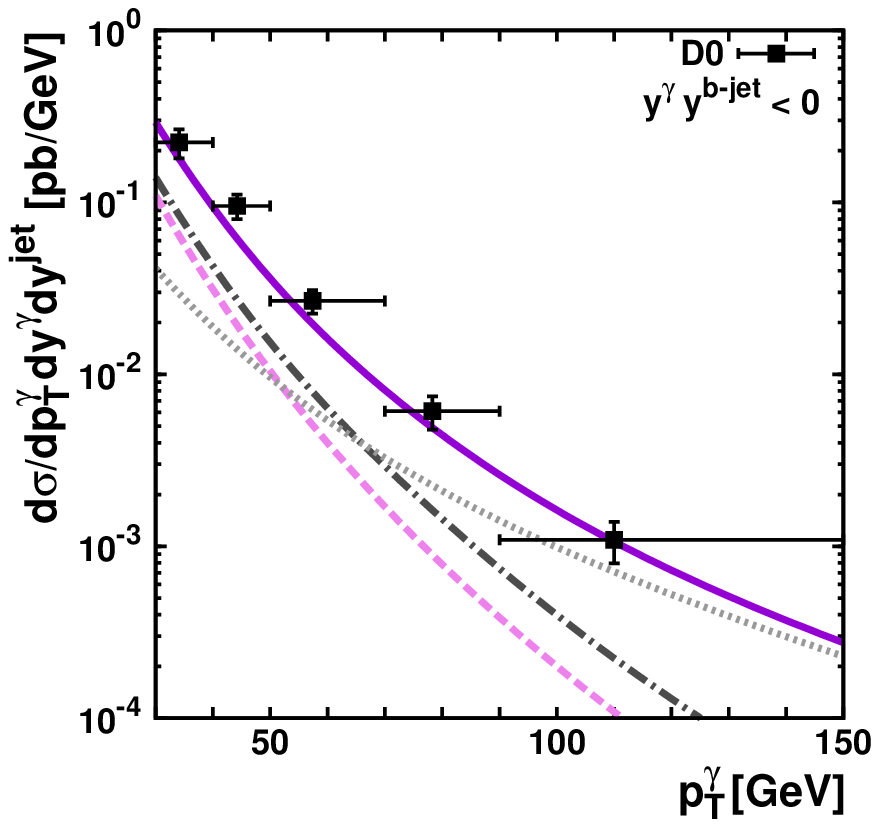, width = 8.1cm}
\caption{The associated $\gamma + b$-jet cross section
as a function of photon transverse momentum $p_T^\gamma$ in the kinematical region defined by
$|y^\gamma| < 1$, $|y^{\rm jet}| < 0.8$ and $p_T^{\rm jet} > 15$~GeV at $\sqrt s = 1960$~GeV. 
Notation of all curves is the same as in Fig.~2. The NLO pQCD predictions\cite{7} are taken 
from\cite{2}. The experimental data are from D$\emptyset$\cite{2}. }
\label{fig3}
\end{center}
\end{figure}

\begin{figure}
\begin{center}
\epsfig{figure=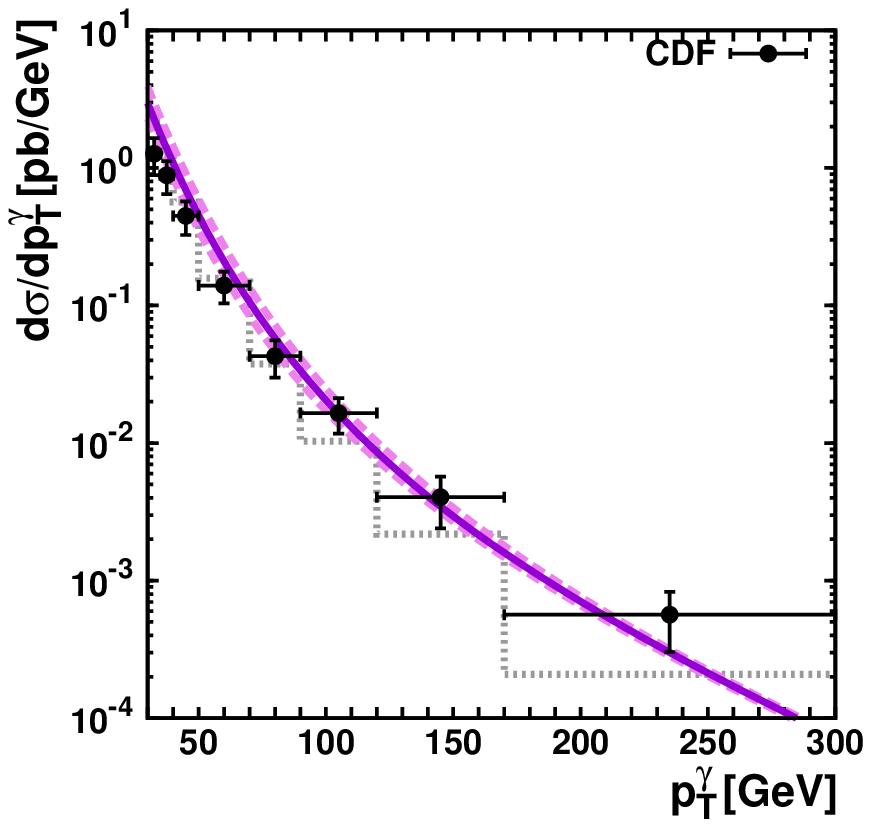, width = 8.1cm}
\epsfig{figure=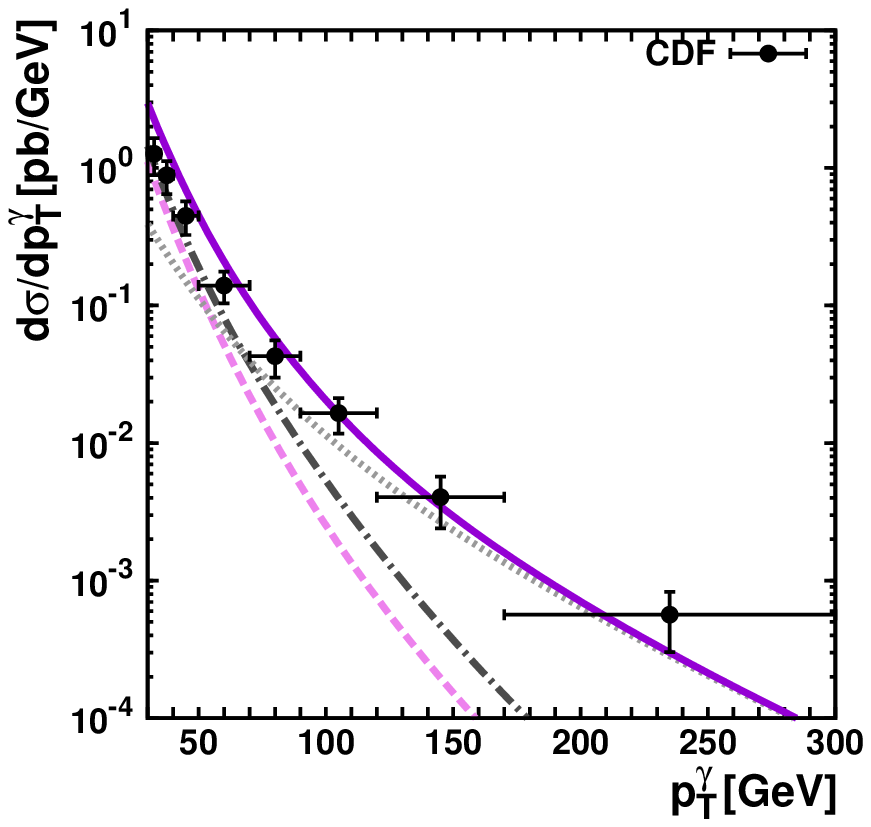, width = 8.1cm}
\caption{The associated $\gamma + b$-jet cross section
as a function of photon transverse momentum $p_T^\gamma$ in the kinematical region defined by
$|y^\gamma| < 1$, $|y^{\rm jet}| < 1.5$ and $p_T^{\rm jet} > 20$~GeV at $\sqrt s = 1960$~GeV. 
Notation of all curves is the same as in Fig.~2. The NLO pQCD predictions\cite{7} are taken 
from\cite{3}. The experimental data are from CDF\cite{3}. }
\label{fig4}
\end{center}
\end{figure}

\begin{figure}
\begin{center}
\epsfig{figure=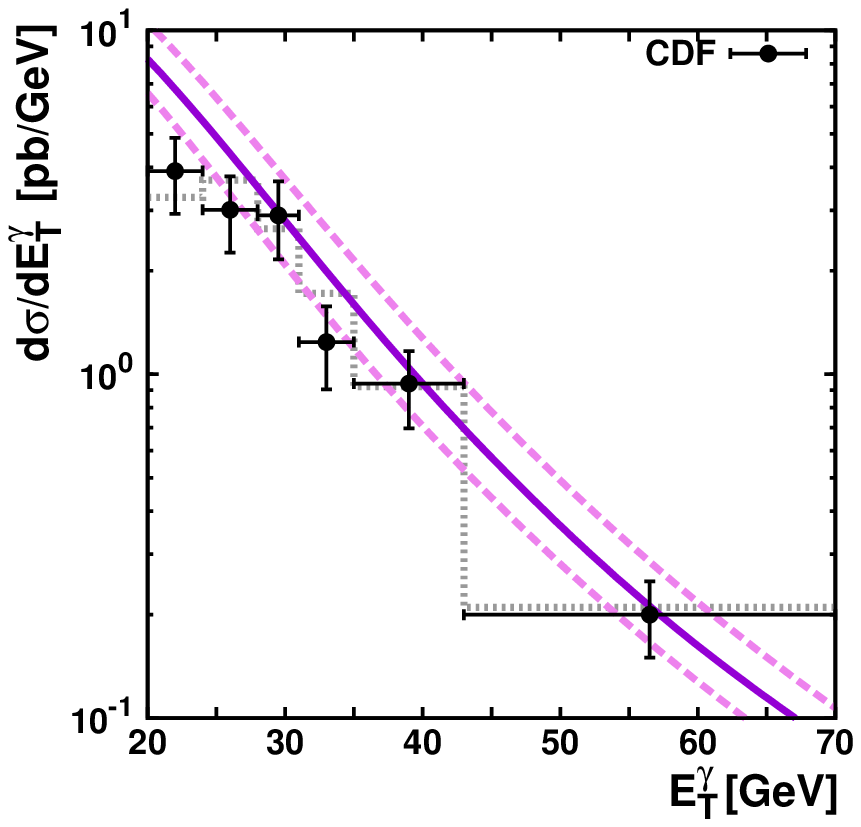, width = 8.1cm}
\epsfig{figure=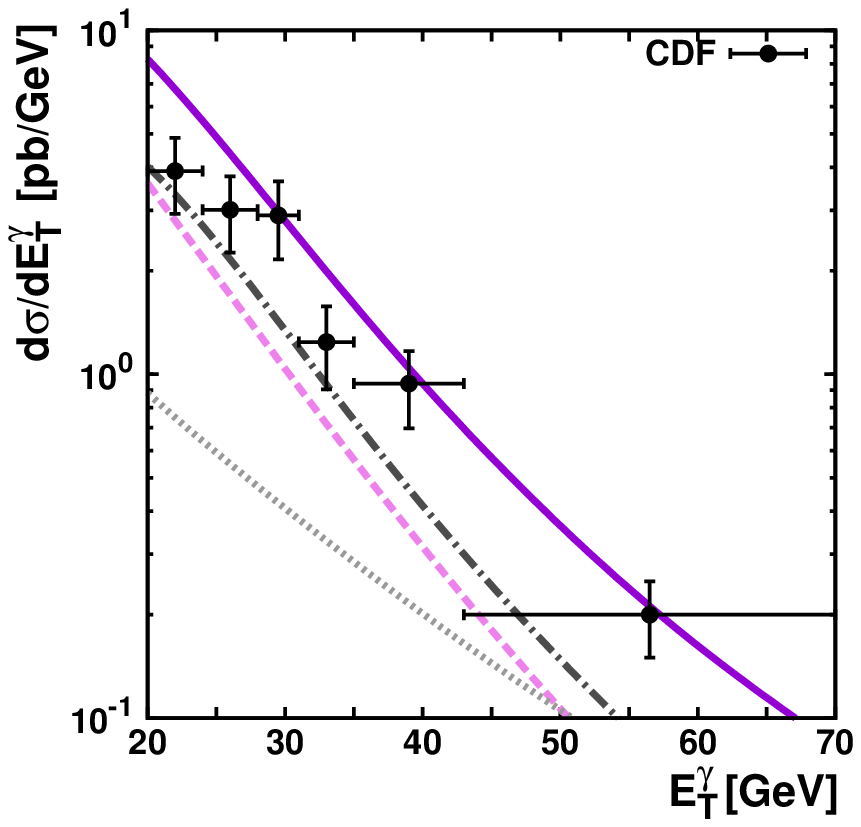, width = 8.1cm}
\caption{The associated $\gamma + b$-jet cross section
as a function of photon transverse energy $E_T^\gamma$ in the kinematical region defined by
$|\eta^\gamma| < 1.1$, $|\eta^{\rm jet}| < 1.5$ and $p_T^{\rm jet} > 20$~GeV at $\sqrt s = 1960$~GeV. 
Notation of all curves is the same as in Fig.~2. The NLO pQCD predictions\cite{7} are taken 
from\cite{4}. The experimental data are from CDF\cite{4}. }
\label{fig5}
\end{center}
\end{figure}

\begin{figure}
\begin{center}
\epsfig{figure=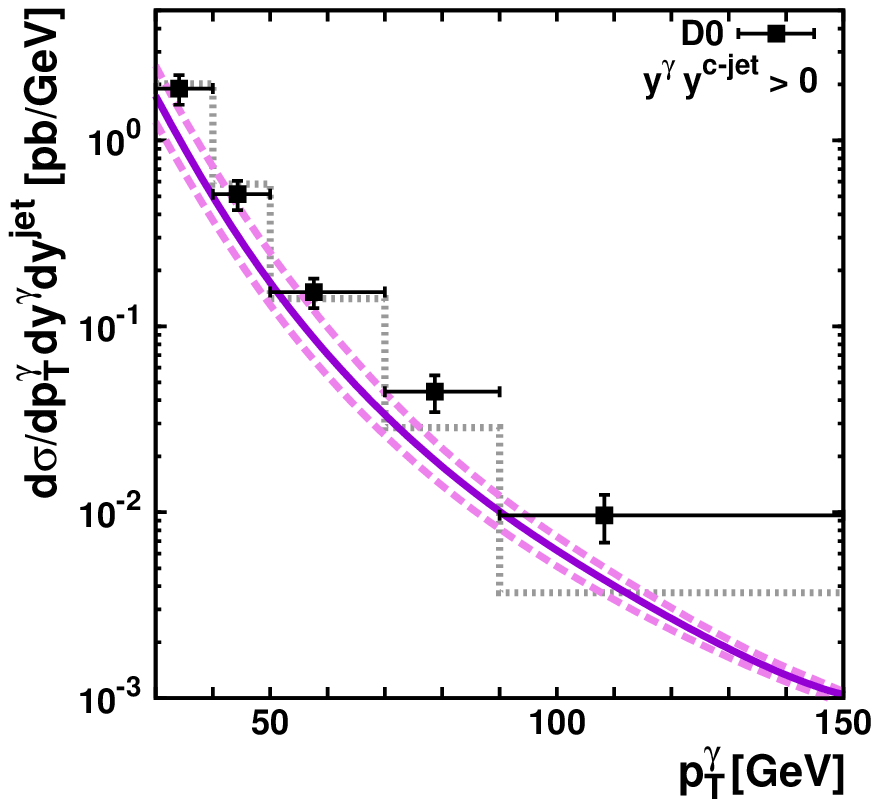, width = 8.1cm}
\epsfig{figure=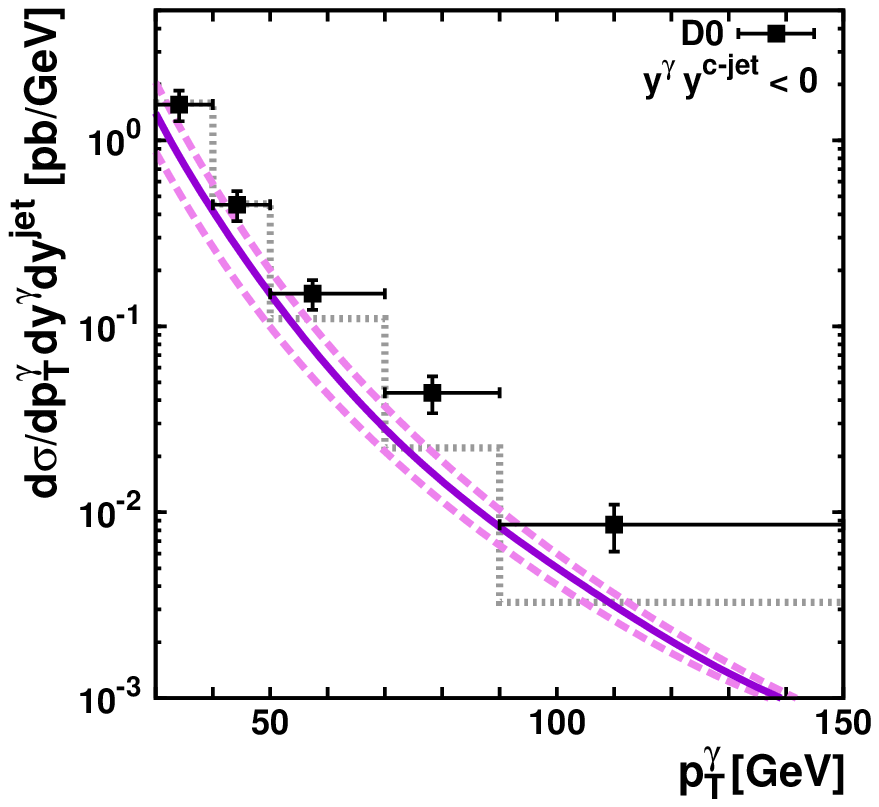, width = 8.1cm}
\caption{The associated $\gamma + c$-jet cross section
as a function of photon transverse momentum $p_T^\gamma$ in the kinematical region defined by
$|y^\gamma| < 1$, $|y^{\rm jet}| < 0.8$ and $p_T^{\rm jet} > 15$~GeV at $\sqrt s = 1960$~GeV. 
Notation of all curves is the same as in Fig.~2. The NLO pQCD predictions\cite{7} are taken
from\cite{7}. The experimental data are from D$\emptyset$\cite{2}. }
\label{fig6}
\end{center}
\end{figure}

\begin{figure}
\begin{center}
\epsfig{figure=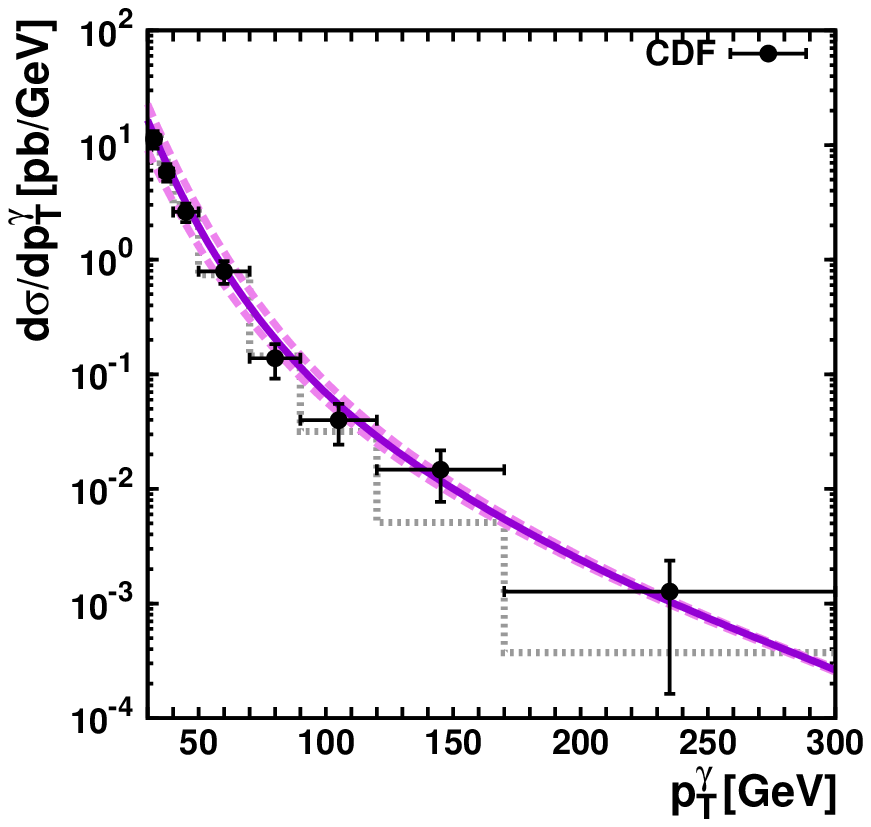, width = 8.1cm}
\epsfig{figure=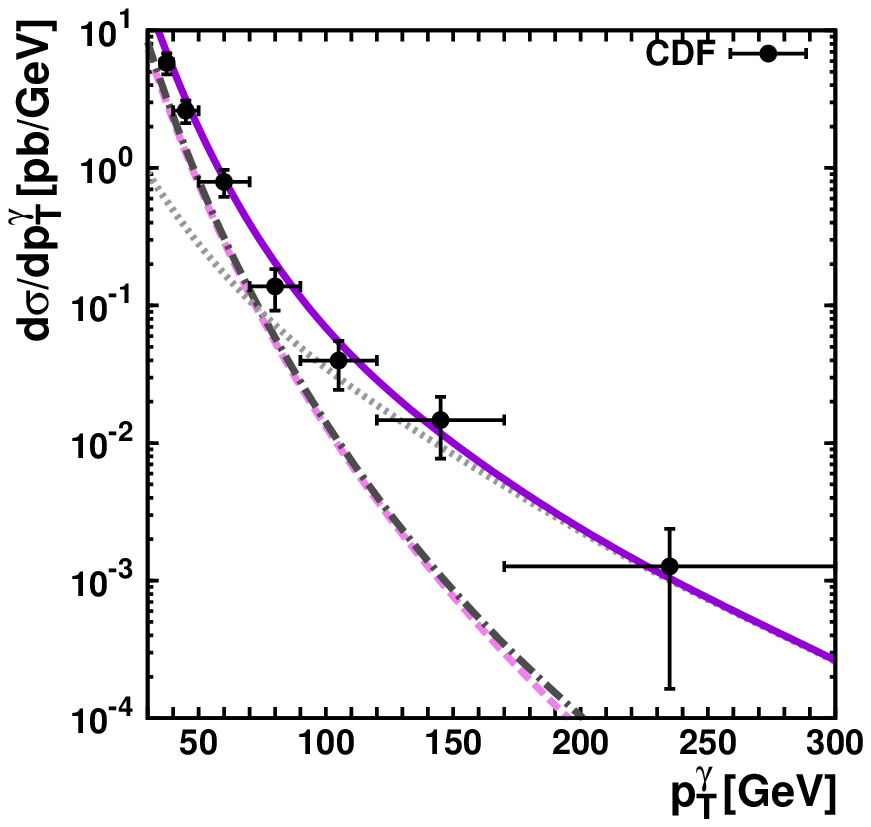, width = 8.1cm}
\caption{The associated $\gamma + c$-jet cross section
as a function of photon transverse momentum $p_T^\gamma$ in the kinematical region defined by
$|y^\gamma| < 1$, $|y^{\rm jet}| < 1.5$ and $p_T^{\rm jet} > 20$~GeV at $\sqrt s = 1960$~GeV. 
Notation of all curves is the same as in Fig.~2. The NLO pQCD predictions\cite{7} are taken 
from\cite{3}. The experimental data are from CDF\cite{3}. }
\label{fig7}
\end{center}
\end{figure}

\begin{figure}
\begin{center}
\epsfig{figure=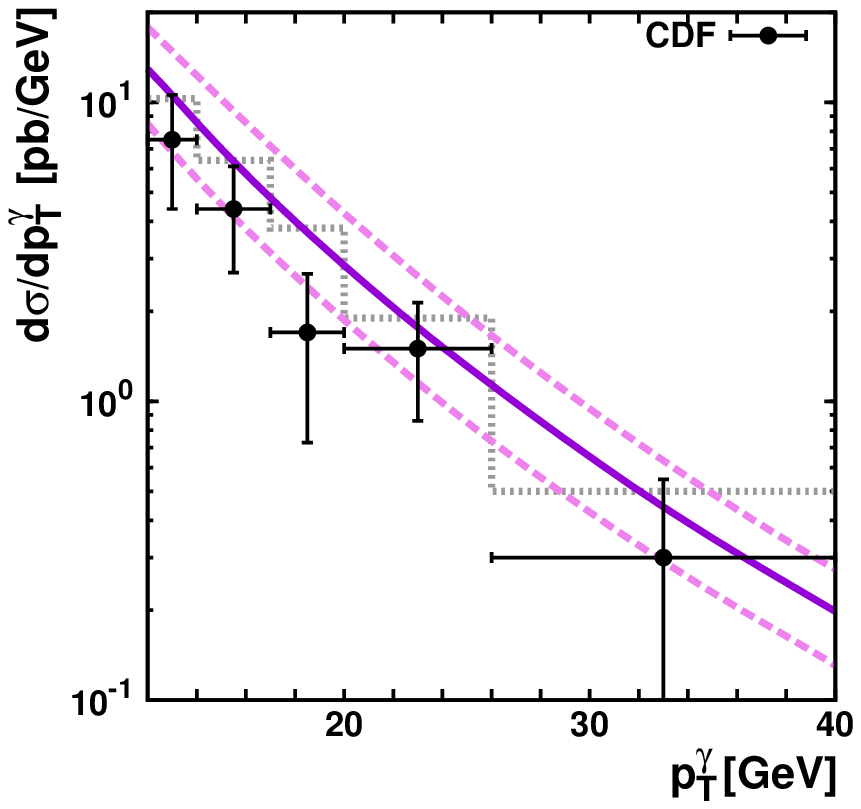, width = 8.1cm}
\epsfig{figure=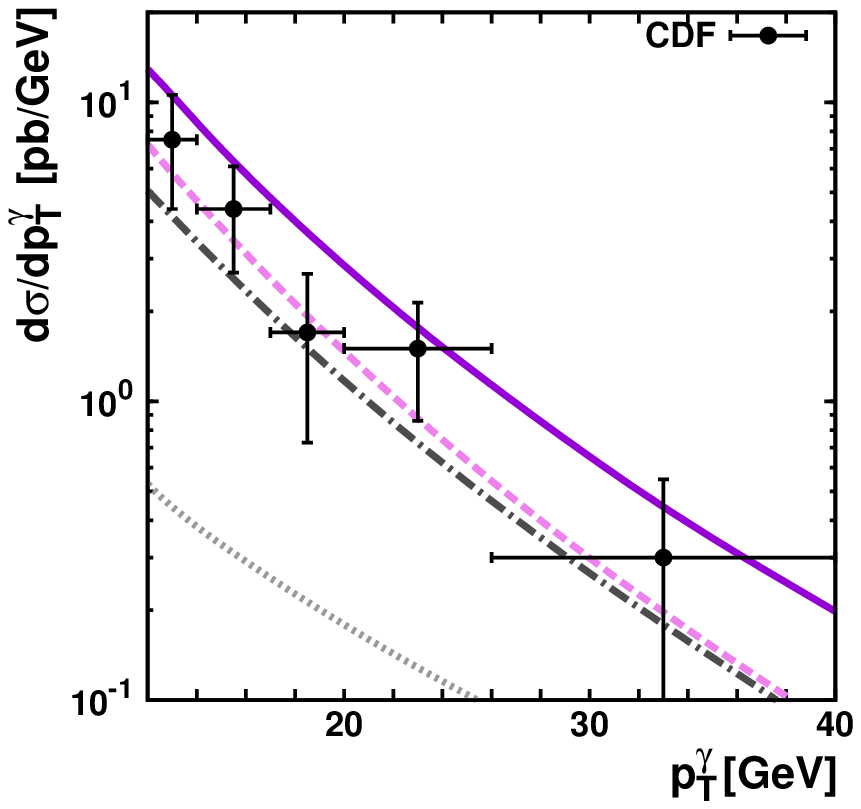, width = 8.1cm}
\caption{The associated $\gamma + \mu$ cross section
as a function of photon transverse momentum $p_T^\gamma$ in the kinematical region defined by
$|\eta^\gamma| < 0.9$, $|\eta^\mu| < 1$ and $p_T^{\mu} > 4$~GeV at $\sqrt s = 1800$~GeV. 
Notation of all curves is the same as in Fig.~2. The NLO pQCD predictions\cite{7} are taken 
from\cite{5}. The experimental data are from CDF\cite{5}. }
\label{fig8}
\end{center}
\end{figure}

\begin{figure}
\begin{center}
\epsfig{figure=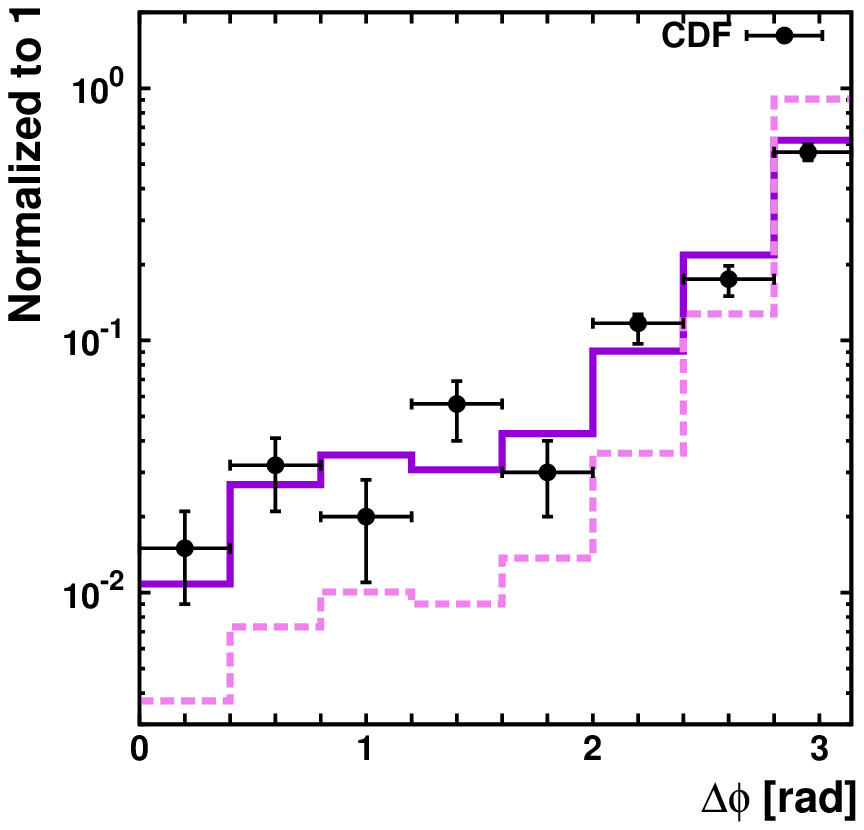, width = 8.1cm}
\epsfig{figure=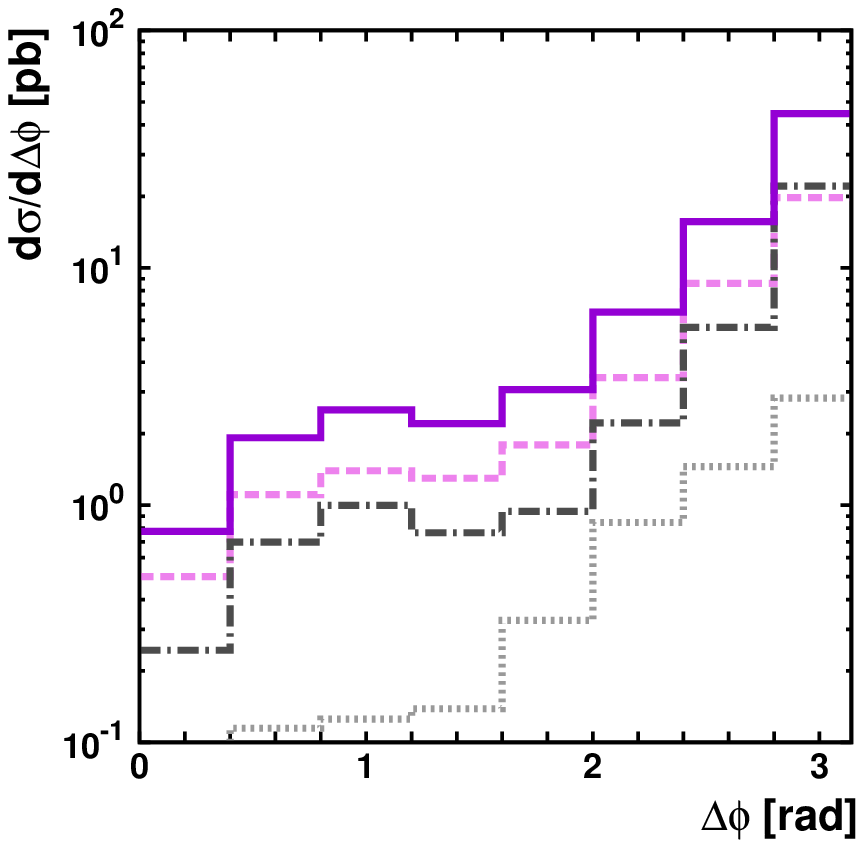, width = 8.1cm}
\caption{The associated $\gamma + \mu$ cross section
as a function of azimuthal angle difference 
$\Delta \phi$ in the kinematical region defined by
$|\eta^\gamma| < 0.9$, $17 < p_T^\gamma < 40$~GeV, $|\eta^\mu| < 1$ and $p_T^{\mu} > 4$~GeV at $\sqrt s = 1800$~GeV. 
Left panel: the solid and dashed curves correspond to the predictions of $k_T$-factorization approach and 
collinear QCD factorization based on the ${\cal O}(\alpha\alpha_s^2)$ matrix elements. 
Notation of all curves in the right panel is the same as in Fig.~2. 
The experimental data are from CDF\cite{6}. }
\label{fig9}
\end{center}
\end{figure}

\begin{figure}
\begin{center}
\epsfig{figure=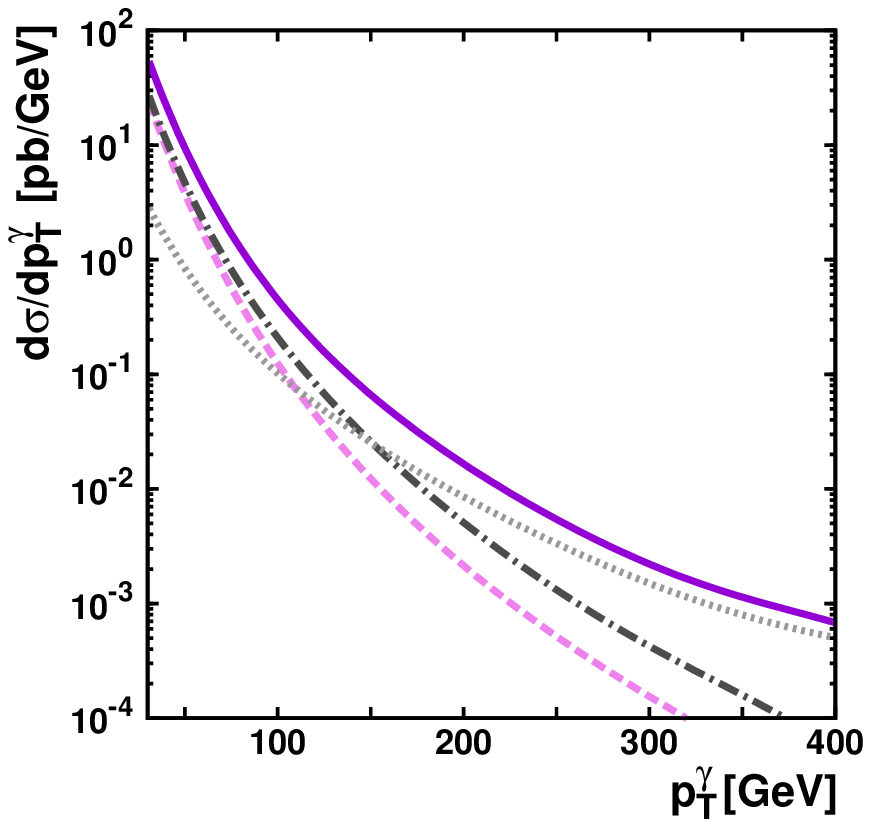, width = 8.1cm}
\epsfig{figure=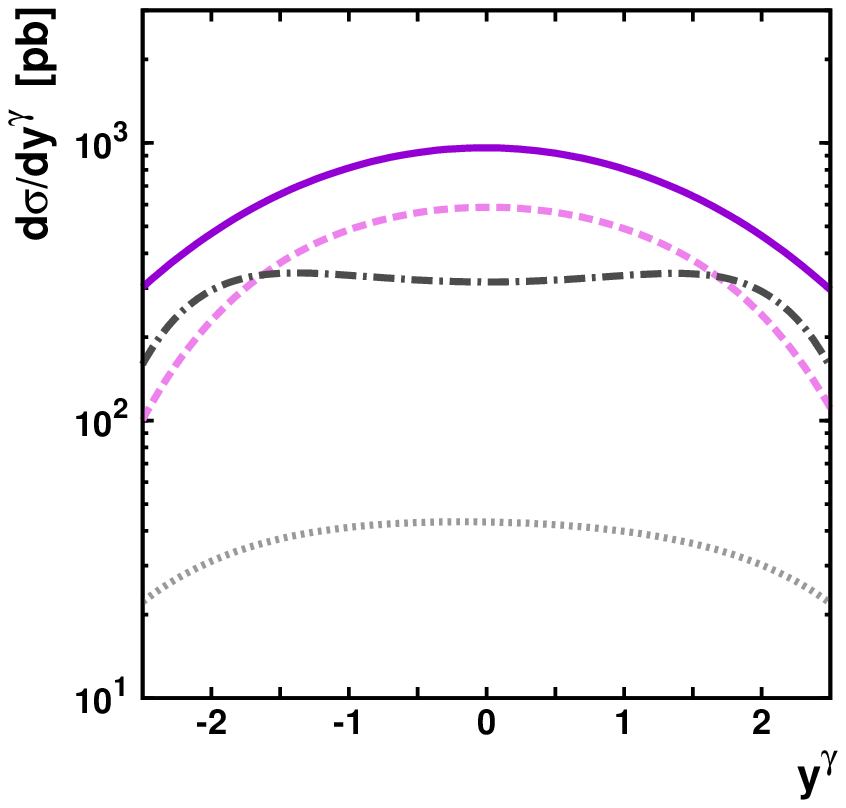, width = 8.1cm}
\caption{The associated $\gamma + b$-jet cross section
as a function of photon transverse momentum $p_{T}^\gamma$ and rapidity
$y^\gamma$ in the kinematical region defined by
$|y^\gamma| < 2.5$, $25 < p_T^\gamma < 400$~GeV, $|y^{\rm jet}| < 2.5$ and $18 < p_T^{\rm jet} < 200$~GeV at $\sqrt s = 7000$~GeV. 
Notation of all curves is the same as in Fig.~2 (right panels).}
\label{fig10}
\end{center}
\end{figure}


\begin{thebibliography}{25}

\bibitem{1} V.M.~Abazov {\sl et al.} (D$\emptyset$ Collaboration), arXiv:1203.5865 [hep-ex]. 
\bibitem{2} V.M.~Abazov {\sl et al.} (D$\emptyset$ Collaboration), Phys. Rev. Lett. {\bf 102}, 192002 (2009). 
\bibitem{3} K.~Vellidis {\sl et al.} (CDF Collaboration), talk given at DIS'12 Workshop, Bonn, March 26 --- 30, 2012. 
\bibitem{4} T.~Aaltonen {\sl et al.} (CDF Collaboration), Phys. Rev. D {\bf 81}, 052006 (2010). 
\bibitem{5} T.~Affolder {\sl et al.} (CDF Collaboration),  Phys. Rev. D {\bf 65}, 012003 (2002). 
\bibitem{6} F.~Abe {\sl et al.} (CDF Collaboration), Phys. Rev. D {\bf 60}, 092003 (1999). 
\bibitem{7} T.P.~Stavreva and J.F.~Owens, Phys. Rev. D {\bf 79}, 054017 (2009). 
\bibitem{8} L.V.~Gribov, E.M.~Levin, and M.G.~Ryskin, Phys. Rep. {\bf 100}, 1 (1983);\\
  E.M.~Levin, M.G.~Ryskin, Yu.M.~Shabelsky and A.G.~Shuvaev, Sov. J. Nucl. Phys. {\bf 53}, 657 (1991);\\
  S.~Catani, M.~Ciafoloni and F.~Hautmann, Nucl. Phys. B {\bf 366}, 135 (1991);\\
  J.C.~Collins and R.K.~Ellis, Nucl. Phys. B {\bf 360}, 3 (1991).
\bibitem{9} H.~Jung, M.~Kr\"amer, A.V.~Lipatov and N.P.~Zotov, JHEP {\bf 1101}, 085 (2011); 
  Phys. Rev. D {\bf 85}, 034035 (2012).
\bibitem{10} S.P.~Baranov, A.V.~Lipatov and N.P.~Zotov, Phys. Rev. D {\bf 81}, 094034 (2010);\\
  A.V.~Lipatov and N.P.~Zotov, Phys. Rev. D {\bf 81}, 094027 (2010); Phys. Rev. D {\bf 72}, 054002 (2005).
\bibitem{11} A.V.~Lipatov and N.P.~Zotov, J. Phys. G {\bf 34}, 219 (2007).
\bibitem{12} S.P.~Baranov, A.V.~Lipatov and N.P.~Zotov, Phys. Rev. D {\bf 77}, 074024 (2008).
\bibitem{13} A.V.~Lipatov, M.A.~Malyshev and N.P.~Zotov, Phys. Lett. B {\bf 699}, 93 (2011).
\bibitem{14} B.~Andersson {\sl et al.} (Small-$x$ Collaboration), Eur. Phys. J. C {\bf 25}, 77 (2002);\\
  J.~Andersen {\sl et al.} (Small-$x$ Collaboration), Eur. Phys. J. C {\bf 35}, 67 (2004);\\
  J.~Andersen {\sl et al.} (Small-$x$ Collaboration), Eur. Phys. J. C {\bf 48}, 53 (2006).
\bibitem{15} T.~Affolder {\sl et al.} (CDF Collaboration), Phys. Rev. D {\bf 65}, 052006 (2002). 
\bibitem{16} S.P.~Baranov, A.V.~Lipatov and N.P.~Zotov, Eur. Phys. J. C {\bf 56}, 371 (2008).
\bibitem{17} J.A.M.~Vermaseren, NIKHEF-00-023.
\bibitem{18} M.A.~Kimber, A.D.~Martin and M.G.~Ryskin, Phys. Rev. D {\bf 63}, 114027 (2001);\\
  G.~Watt, A.D.~Martin and M.G.~Ryskin, Eur. Phys. J. C {\bf 31}, 73 (2003).
\bibitem{19} A.D.~Martin, W.J.~Stirling, R.S.~Thorne and G.~Watt, Eur. Phys. J. C {\bf 63}, 189 (2009).
\bibitem{20} G.P.~Lepage, J. Comput. Phys. {\bf 27}, 192 (1978).
\bibitem{21} J.~Pumplin, H.L.~Lai and W.K.~Tung, Phys. Rev. D {\bf 75}, 054029 (2007).
\bibitem{22} C.~Peterson, D.~Schlatter, I.~Schmitt and P.~Zerwas, Phys. Rev. D {\bf 27}, 105 (1983).
\bibitem{23} K.~Nakamura {\sl et al.} (PDG Collaboration), J. Phys. G {\bf 37}, 075021 (2010).
\bibitem{24} CMS Collaboration, Phys. Rev. D {\bf 84}, 052011 (2011); arXiv:1202.4617 [hep-ex].
\bibitem{25} ATLAS Collaboration, Phys. Lett. B {\bf 706}, 150 (2011).

\end{thebibliography}
\end{document}